# First-principles study of electronic and magnetic properties of self-intercalated van der Waals magnet $Cr_3Ge_2Te_6$

Jia-wan Li(李家万), Shi-Bo Zhao(赵世博), Lin Zhuang(庄琳),

Yusheng Hou(侯玉升)[†]

Guangdong Provincial Key Laboratory of Magnetoelectric Physics and Devices,
Center for Neutron Science and Technology, School of Physics, Sun Yat-Sen,
Guangzhou, 510275, China

Self-intercalated van der Waals magnets, characterized by self-intercalating native atoms into van der Waals layered structures with intrinsic magnetism, exhibit a variety of novel physical properties. Here, using first-principles calculations and Monte Carlo simulations, we report a self-intercalated van der Waals ferromagnet, $Cr_3Ge_2Te_6$, which has a high Curie temperature of 492 K. We find that $Cr_3Ge_2Te_6$ is nearly half-metallic with a spin polarization reaching up to 90.9%. Due to the ferromagnetism and strong spin-orbit coupling effect in $Cr_3Ge_2Te_6$, a large anomalous Hall conductivity of $138\ \Omega^{-1}cm^{-1}$ and $305\ \Omega^{-1}cm^{-1}$ can be realized when its magnetization is along its magnetic easy axis and hard axis, respectively. By doping electrons (holes) into $Cr_3Ge_2Te_6$, these anomalous Hall conductivities can be increased up to $318\ \Omega^{-1}cm^{-1}$ ($648\ \Omega^{-1}cm^{-1}$). Interestingly, a 5-layer $Cr_3Ge_2Te_6$ thin film retains the room-temperature ferromagnetism with a higher spin polarization and larger anomalous Hall conductivity. Our work demonstrates that $Cr_3Ge_2Te_6$ is a novel room-temperature self-intercalated ferromagnet with high spin polarization and large anomalous Hall conductivity, offering great opportunities for designing nano-scale electronic devices.

[†]Corresponding authors: houysh@mail.sysu.edu.cn

# 1. Introduction

Since atomically thin graphene was mechanically exfoliated from conventional van der Waals (vdW) layered graphite in 2004 [1], the research of vdW layered materials has triggered significant interest [2]. Especially, magnetic vdW materials featuring intrinsic magnetism provide an additional degree of freedom, i.e., the spin of the electron, which potentially offers greater opportunity for manipulating electronic states [3, 4]. For instance, recent studies have observed intriguing phenomena such as the giant magneto-optical effect in vdW room-temperature ferromagnet $Fe_3GaTe_2$ and large room-temperature magnetoresistance in its heterostructures [5, 6]. Furthermore, the coexistence of cleavable properties and magnetism makes vdW magnets an ideal platform for exploring magnetism in the two-dimensional (2D) limit [7]. Experimental discoveries of FM order in the 2D thin film of Cr-based vdW magnets $CrI_3$ [8] and $Cr_2Ge_2Te_6$ [9] underscore their potential. However, despite extensive research, the low magnetic critical temperature ($T_C$) in these 2D vdW magnets hinders their practical applications in spintronic devices.

By intercalating foreign guest species, e.g., lithium and tetrabutyl ammonium, into the vdW gap of these layered magnets, their $T_C$s can be greatly enhanced and their practical applications in spintronic devices is thereby enabled [10, 11]. However, the ill-defined intercalated phase in such intercalated vdW magnets renders their theoretical studies difficult. Intercalating native atoms into transition metal dichalcogenides (TMDs), a new class of self-intercalated vdW magnets with the advantages of high-quality products, well-defined phases and long-range crystalline order have received more and more attention [12]. Among them, a series of non-ferromagnetic (Ti-, Co-, Nb-, Mo-, Ta-based) TMDs develop FM order upon self-intercalation [13]. In the case of those with intrinsic magnetism, such as FM $CrTe_2$, different stoichiometric self-intercalated $Cr_{1+\delta}Te_2$ ($0 < \delta < 1$) phases are produced by intercalating Cr atoms under growth. These novel phases can retain the ferromagnetism and induce novel magnetic and topological properties [14]. For example, strong magnetic anisotropy and colossal anomalous Hall effect (AHE) have been observed in $Cr_5Te_8$ [15-17]. A giant topological hall effect has been noted in room-temperature ferromagnet $Cr_{0.82}Te$ [18]. Additionally, a topological Hall effect, along with potential magnetic skyrmion phase, has been identified in $Cr_2Te_2$ [19]. A Néel-type skyrmion phase below 200 K has been observed in acentric self-intercalated $Cr_2Te_3$ [20]. Thickness-dependent magnetic biskyrmions have been found in $Cr_3Te_4$ nanosheet [21]. Lastly, room-temperature magnetic skyrmions have been reported in self-intercalated $Cr_{1.53}Te_2$ [22].

Recently, another self-intercalated vdW magnet with a ferrimagnetic (FiM) order, $Mn_3Si_2Te_6$, has also drawn intensive research interest for its colossal magnetoresistance (MR) [23-33]. One of the most inspiring characteristics is its spin orientation-dependent MR, which changes by 7 orders of magnitude when a magnetic field is applied along its magnetic hard axis [23, 24]. When $Mn_3Si_2Te_6$ is thinned down to 2D limit, its nanoflakes can exhibit an electrically gate-tunable MR with a faster response [31]. In addition, an unconventional chiral orbital current state flowing along the edges of $MnTe_6$ octahedra enables the current-sensitive giant Hall effect in $Mn_3Si_2Te_6$ [26, 34]. Theoretically, switchable in-plane AHE by magnetization orientation is put forward in $Mn_3Si_2Te_6$ monolayer [35]. Other merits in $Mn_3Si_2Te_6$, such as magnetocaloric effect [36], low thermal conductivity [37, 38], and anomalous Nernst effect [39], imply its potential applications in thermoelectricity. The enhanced ultrafast terahertz conductivity [40] and large magneto-optical effects [41] highlight the photoelectric applications of $Mn_3Si_2Te_6$. Considering that the low $T_C$ (~78 K) of $Mn_3Si_2Te_6$ [42] hinders its practical applications of these abundant exotic properties, some strategies, including high pressure [27, 29, 32, 33] and electron doping [43], are proposed to boost its $T_C$. However, the $T_C$ of $Mn_3Si_2Te_6$ in these studies is still lower than room temperature, highlighting the need to find new self-intercalated vdW magnets.

In this paper, we systematically investigate the electronic and magnetic properties of a novel self-intercalated vdW magnet $Cr_3Ge_2Te_6$, a sibling of $Mn_3Si_2Te_6$, using first-principles calculations and Monte Carlo simulations. Unlike $Mn_3Si_2Te_6$, which is FiM and has a low $T_C$, our DFT calculations reveal that $Cr_3Ge_2Te_6$ is thermodynamically stable and an in-plane ferromagnet with a high $T_C$ of 492K which is a prerequisite for practical applications. By examining its band structures, we find that $Cr_3Ge_2Te_6$ is nearly half-metallic with a spin polarization of 90.9%. As a result of the ferromagnetism and the strong spin-orbit coupling (SOC) effect, $Cr_3Ge_2Te_6$ exihibits an anomalous Hall conductivity (AHC) of 138 $\Omega^{-1}cm^{-1}$. When its magnetization is regulated along the hard axis (i.e., $z$ axis), $Cr_3Ge_2Te_6$ has a larger AHC of 305 $\Omega^{-1}cm^{-1}$. Because the AHC is sensitive to the Fermi level, it can be further enhanced by doping electrons or holes into $Cr_3Ge_2Te_6$. Lastly, we show that a 5-layer (5L) $Cr_3Ge_2Te_6$ thin film also possesses in-plane ferromagnetism with a high $T_C$ above room-temperature and a higher spin polarization of 99.8%. When the 5L $Cr_3Ge_2Te_6$ thin film has an out-of-plane magnetization, its AHC can be as large as 318 $\Omega^{-1}cm^{-1}$ and further increased up to 731 $\Omega^{-1}cm^{-1}$ through doping holes. It is worth noting that the high $T_C$, high spin polarizatioin and out-of-plane AHE in the 5L $Cr_3Ge_2Te_6$ thin film is different from the magnetic properties and in-plane AHE in the $Mn_3Si_2Te_6$ thin film [31, 35]. Our work demonstrates that the room-temperature ferromagnetism, high spin polarization and large AHC in vdW self-intercalated magnet $Cr_3Ge_2Te_6$ may provide opportunities for developing intriguing spintronic devices.

## 2. Computational methods

Our first-principles calculations based on the density-functional theory (DFT) are performed using the Vienna *ab initio* simulation package (VASP) [44] at the level of the generalized gradient approximation with Perdew-Burke-Ernzerhof potential [45]. The pseudopotentials with valence electrons of $Cr3d^54s^1$, $Ge4s^24p^2$ and $Te5s^25p^4$ are utilized through the projector augmented wave method [46, 47]. The cutoff energy is 350 eV and the total energy criterion at the self-consistent step is set to be $10^{-6}$ eV. In structural relaxations, the lattice constants and atomic positions are fully relaxed until the force acting on each atom is smaller than 0.01 eV/Å. Furthermore, the local-spin-density approximations plus $U$ (LSDA+$U$) method [48] and the effective Coulomb interaction $U_{eff}$ = 2.0 eV with $U$=3.0 eV and $J_H$= 1.0 eV are adopted to describe the strong correlation effects among $3d$ electrons of Cr [49]. Based on the magnetic parameters obtained from DFT calculations, we perform Monte Carlo simulations using the Metropolis algorithm to explore magnetic ground states and $T_C$. The 5L $Cr_3Ge_2Te_6$ thin film is built with a vacuum space of 15 Å to avoid spurious interactions.

## 3. Results and discussions

Figure 1(a) shows the self-intercalated vdW structure of $Cr_3Ge_2Te_6$, which consists of two different Cr-sites (denoted as Cr1 and Cr2, respectively). The layers formed by Cr1 atoms, where $CrTe_6$ octahedra share edges within the *ab* plane, are analogous to the honeycomb structure of $Cr_2Ge_2Te_6$. Cr2 atoms are intercalated between two Cr1 layers, and their half-filled structures in the vdW gap form a Cr2-trigonal layer. Thus, as shown in Fig. 1(b), Cr1-honeycomb and Cr2-trigonal layers alternatively stack along the *c* axis, giving rise to the same trigonal space group $P\bar{3}1c$ as $Mn_3Si_2Te_6$ [42]. Our calculations show that the lattice constants of $Cr_3Ge_2Te_6$ are $a = b = 7.12$ Å and $c = 14.52$ Å, which are comparable to those in $Mn_3Si_2Te_6$ ($a = b = 7.03$ Å, $c = 14.25$ Å). This is understandable because Cr atom with a $3d^54s^1$ electronic configuration has the similar radius as Mn ($3d^54s^2$) [42]. In $Cr_3Ge_2Te_6$, the Ge-Ge bond length along the *c* axis is 2.47 Å, comparable with that in $Cr_2Ge_2Te_6$ [50]. This indicates that Ge atoms form Ge-Ge dimers. As a result, Ge atoms have a valence state of 3+. Besides, Te atoms exhibit commonly a valence state of 2- due to their strong electronegativity. Consequently, Cr atoms are in a formal 2+ valence state and have a $3d^4$ configuration to maintain the electrical neutrality in $Cr_3Ge_2Te_6$. Our calculated magnetic moments for Cr1 and Cr2 atoms are 3.86 and 4.06 $\mu_B$/Cr, respectively. These indicate that $Cr^{2+}$ cations have a high-spin state with a spin quantum number of 2.

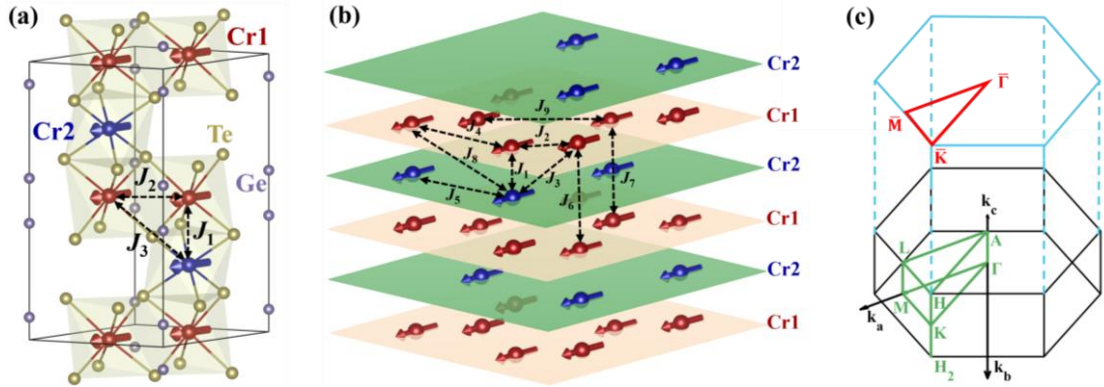

**Fig. 1.** (a) Side view of the self-intercalated structure of $Cr_3Ge_2Te_6$. $CrTe_6$ octahedra are shown by the dark yellow color. Double-head arrows highlight the first three nearest-neighbor exchange paths. (b) Alternatingly stacked structure with only Cr atoms being shown. Cr1-honeycomb and Cr2-trigonal layers are indicated by light pink and green parallelograms, respectively. Double-head arrows show the first nine nearest-neighbor exchange paths. (c) The first Brillouin zone. Black lines and light blue lines represent the Brillouin zone of bulk and thin film, respectively. The high symmetry paths with labels for bulk and thin film are shown by green and red, respectively.

To confirm the thermodynamical stability of $Cr_3Ge_2Te_6$, we calculate its phonon spectrum with finite displacement method [51] and perform *ab* initio molecular dynamic simulations at 300K with the NVT canonical ensemble and Nosé–Hoover thermostat [52]. Note that the magnetic ground state of the FM order (see the discussions below) is adopted in the phonon spectrum calculation. Figure 2(a) depicts the phonon dispersions along the high symmetry paths shown in Fig. 1(c). One can see that there is no imaginary frequency in $Cr_3Ge_2Te_6$, suggesting that it is dynamically stable. Figure 2(b) shows the evolution of energy and temperature over time in *ab* initio molecular dynamic simulations. We see that the energy and temperature fluctuate around the equilibrium level. In addition, the crystal structure of $Cr_3Ge_2Te_6$ is not broken down in our simulation (see the insets in Fig. 2(b)). All of these indicate that $Cr_3Ge_2Te_6$ is thermally stable at room temperature. Therefore, our calculations confirm that $Cr_3Ge_2Te_6$ is thermodynamically stable and could be synthesized in future experiments.

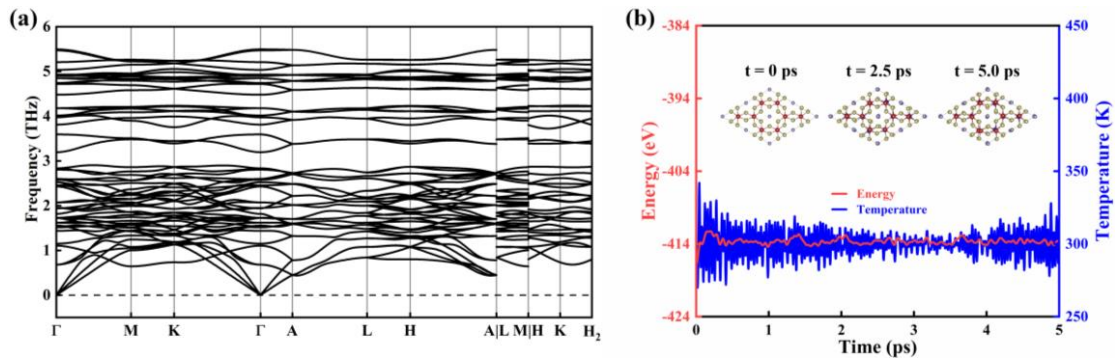

**Fig. 2.** Thermodynamical stability of $Cr_3Ge_2Te_6$. (a) Phonon spectrum along the high

symmetry paths. (b) The time-dependent evolutions of energy and temperature under the temperature of 300 K. The insets show the top views of the crystal structures of $Cr_3Ge_2Te_6$ at 0, 2.5 and 5.0 ps, respectively.

Here, we first investigate the magnetic properties of $Cr_3Ge_2Te_6$. To determine its magnetic ground state, we consider four typical magnetic orders of self-intercalated vdW structure. These orders are FM, FiM and two antiferromagnetic (denoted as AFM1 and AFM2, respectively) orders, which are depicted in Fig. 3(a)-(d). Note that the AFM1 and AFM2 orders are two different antiferromagnetism with fully compensated magnetizations. As shown in Fig. 3(e), the calculated total energies of these four magnetic orders show that the FM order has the lowest total energy. Among them, the FiM and AFM1 orders have a significantly higher total energy than the FM order. Explicitly, their total energies are higher than that of the FM order by 152 and 198 meV/Cr. In short, our DFT calculated results suggest that the FM order is the magnetic ground state of $Cr_3Ge_2Te_6$.

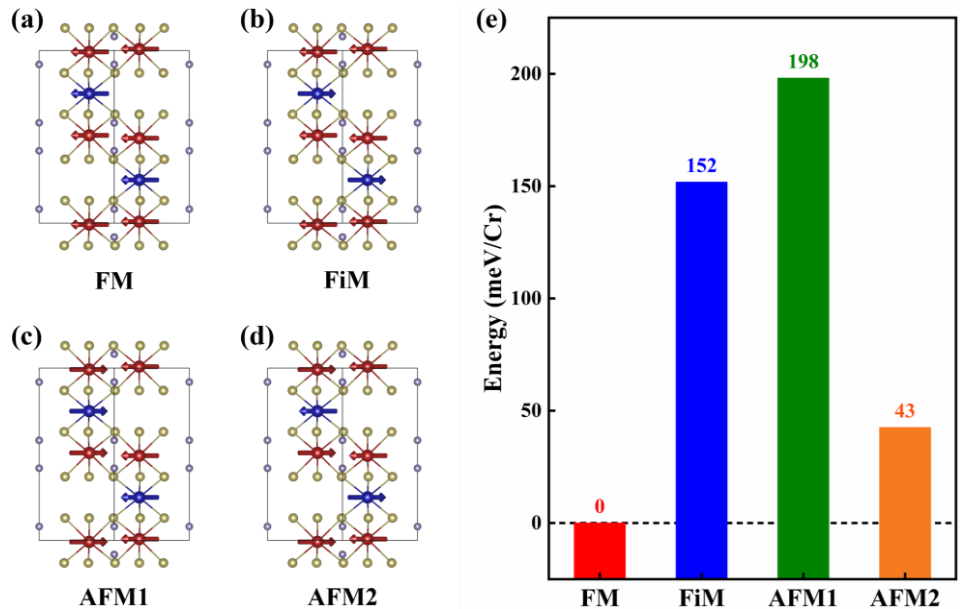

**Fig. 3.** The schematic of the considered four magnetic orders and DFT calculated total energies. (a) FM, (b) FiM, (c) AFM1 and (d) AFM2 orders. (e) DFT calculated total energies of the four magnetic orders. The total energy of FM order is set as reference.

We employ a spin Hamiltonian that is composed of the first nine nearest-neighbor (NN) Heisenberg exchange interactions [see Fig. 1(b)] and magnetic anisotropy energy (MAE) to determine the $T_C$ of $Cr_3Ge_2Te_6$. Our spin Hamiltonian has the following form,

$$H_{spin} = \sum_{i<j} J_{ij} \vec{S_i} \cdot \vec{S_j} + K \sum_i (S_i^z)^2 \qquad (1)$$

In Eq. (1), $\vec{S_i}$ and $\vec{S_j}$ are the spins at the magnetic sites $i$ and $j$, respectively; the spin $S_i^z$ is defined by the local $z$ axis of each ion $i$; $J_{ij}$ is the Heisenberg exchange interactions, and $K$ is the MAE constant. Antiferromagnetic (AFM) and FM Heisenberg interaction are characterized by a positive and negative $J_{ij}$, respectively. $K$ is related to the energy difference ($E_z - E_x$) between spin orientations of $z$ ($E_z$) and $x$ ($E_x$) axes, with positive (negative) values of $K$ representing in-plane (out-of-plane) magnetic anisotropy. As listed in Table 1, our calculations demonstrate that the values of Heisenberg exchange couplings are almost negative. Only three of them are relatively small positive, indicating that $Cr_3Ge_2Te_6$ is dominated by FM interactions. The first three NN Heisenberg exchange couplings are all FM, which is different from the strong magnetic frustration in $Mn_3Si_2Te_6$ [42]. Within the Cr1-honeycomb layer, FM $J_2$ and $J_9$ are much stronger than AFM $J_4$. Thus, a FM order is formed in this layer. Although there is an AFM $J_5$ (6.44 meV) in the Cr2-trigonal layers, the FM $J_3$ (-41.60 meV) between Cr1-honeycomb layers and Cr2-trigonal layers is very strong. Consequently, a FM order is also formed in the Cr2-trigonal layer. The positive MAE (0.55 meV/Cr) indicates that $Cr_3Ge_2Te_6$ has an in-plane magnetic easy axis (i.e., $x$ axis). Lastly, our Monte Carlo simulations confirm that $Cr_3Ge_2Te_6$ has an in-plane FM magnetic ground state, with a $T_C$ as high as 492 K. Hence, $Cr_3Ge_2Te_6$ is a room-temperature ferromagnet.

**Table 1.** Magnetic parameters of $Cr_3Ge_2Te_6$ bulk and its 5L thin film. Heisenberg exchange interactions between two layers (i.e., Cr1-honeycomb and Cr2-trigonal layers) are classified in the column with header "Type". The distances and values of nine NN exchange paths are in units of Å and meV, respectively. The MAE constant $K$ and $T_C$ are in units of meV/Cr and K, respectively. In 5L $Cr_3Ge_2Te_6$ thin film, distances of the first, third and eighth exchange paths (i.e., $J_1$, $J_3$ and $J_8$) are split into two slightly different values, respectively.

| Exchange paths | Type | Bulk | | 5L | |
| --- | --- | --- | --- | --- | --- |
| | | Distance | Value | Distance | Value |
| $J_1$ | Cr1-Cr2 | 3.603 | -0.76 | 3.729 | -7.50 |
| | | | | 3.731 | -11.45 |
| $J_2$ | Cr1-Cr1 | 4.105 | -9.22 | 4.073 | -10.58 |
| $J_3$ | Cr1-Cr2 | 5.398 | -41.60 | 5.471 | -40.91 |
| | | | | 5.516 | -38.36 |
| $J_4$ | Cr1-Cr1 | 7.013 | 1.33 | 7.054 | 0.63 |
| $J_5$ | Cr2-Cr2 | 7.109 | 6.44 | 7.054 | 1.58 |
| $J_6$ | Cr1-Cr1 | 7.109 | 8.67 | 7.374 | 5.27 |
| $J_7$ | Cr1-Cr1 | 7.206 | -3.11 | 7.460 | -5.69 |
| $J_8$ | Cr1-Cr2 | 7.970 | -0.40 | 7.979 | 0.71 |
| | | | | 7.980 | 0.16 |
| $J_9$ | Cr1-Cr1 | 8.209 | -9.30 | 8.145 | -8.16 |
| $K$ | | 0.55 | | 0.25 | |
| $T_C$ | | 492 | | 409 | |
| Magnetic ground state | | FM | | FM | |

Now, we study the electronic properties of the FM $Cr_3Ge_2Te_6$. Our spin-resolved band structure, depicted in Fig. 4(a), shows that $Cr_3Ge_2Te_6$ is definitely a metal. We find that the bands passing through the Fermi level are almost contributed by the majority states (i.e., spin up), while the conduction bands of minority states (i.e., spin down) around the high symmetry points of K and H slightly across the Fermi level. These features indicate that $Cr_3Ge_2Te_6$ is nearly half-metallic. To quantify this nearly half-metallicity in $Cr_3Ge_2Te_6$, we adopt the most commonly used definition of spin polarization ($P$) as follows [53],

$$P = \frac{N_\uparrow - N_\downarrow}{N_\uparrow + N_\downarrow} \qquad (2).$$

In Eq. (2), $N_\uparrow$ and $N_\downarrow$ are the total density of states (DOS) of majority and minority states at the Fermi level, respectively. For simplicity, the integral total DOS of 18 meV near the Fermi level [see the cyan area in Fig. 4(b)] from our DFT calculations is used to approximate the $P$. Based on Eq. (2), we obtain that $Cr_3Ge_2Te_6$ is nearly a half-metal with a high spin polarization of up to 90.9%. Additionally, atom-resolved DOS reveals that the bands around the Fermi level mainly come from Cr and Te atoms, implying the strong *p-d* hybridizations between them [see Fig. 4(b)]. Figure 4(c) shows that the 3*d* orbitals of $Cr^{2+}$ cations split into three distinct levels, consistent with the point group $D_{3d}$ in $Cr_3Ge_2Te_6$. For a $Cr^{2+}$ cation, its four 3*d* electrons are in high-spin state and lead to 4 $\mu_B$/Cr, aligning with our calculated magnetic moments. Considering the effect of strong SOC in Te atoms on electronic properties, we then examine the band structures of the FM $Cr_3Ge_2Te_6$ with SOC included in our DFT calculations. To obtain a deeper understanding on the effect of SOC, we set the magnetization of $Cr_3Ge_2Te_6$ along the *x* ($M \parallel x$) and *z* ($M \parallel z$) axes, respectively. As shown in Fig. 4(c) and (d), some crossing points and two-fold degenerate lines of bands are split. For instance, anti-crossing points along K-Γ (Γ-M) and gapped lines along A-L (Γ-A) occur in the case of $M \parallel x$ ($M \parallel z$). Thus, combining the ferromagnetism with a strong SOC effect, some important properties, such as AHE, may be expected in $Cr_3Ge_2Te_6$.

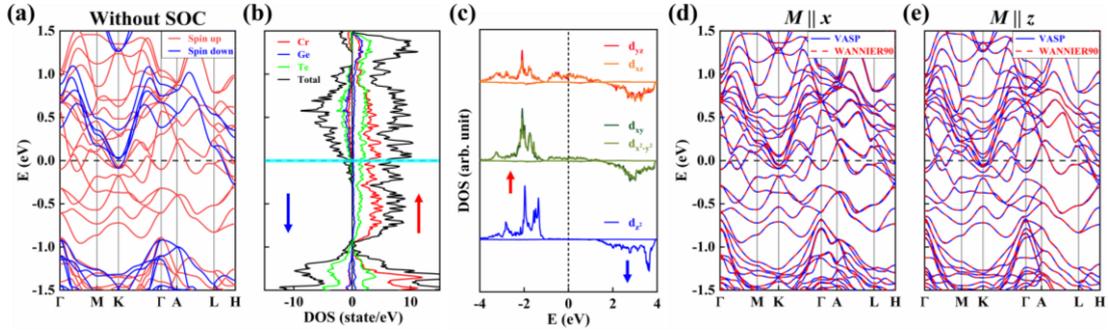

**Fig. 4.** Electronic properties of $Cr_3Ge_2Te_6$. (a) Spin-resolved band structure without SOC. (b) Atom-resolved DOS. The cyan energy interval around the Fermi level is used to integrate the total DOS. (c) The 3*d*-orbital-resolved DOS. (d) Band structure in the case of $M \parallel x$ when SOC is considered. (e) same as (d) but for the case of $M \parallel z$. In (d) and (e), DFT-calculated and Wannier-interpolated bands are represented by blue solid and red dash lines, respectively.

To investigate the underlying AHE in the FM $Cr_3Ge_2Te_6$, we construct an effective tight binding model with WANNIER90 package [54]. As shown in Fig. 4(c) and (d), the Wannier-interpolated bands fit well with those of DFT calculations. We then use

WannierTools package [55] to calculate the dependence of AHC on the position of the Fermi level of $Cr_3Ge_2Te_6$. As shown in Fig. 5(a), when the magnetization is along the easy axis (i.e., $M \parallel x$), the AHC component $\sigma_{yz}$ of $Cr_3Ge_2Te_6$ is basically within 500 $\Omega^{-1}cm^{-1}$. Given the weak MAE (0.55 meV/Cr) in $Cr_3Ge_2Te_6$, the magnetization can be easily regulated to the hard axis (i.e., $M \parallel z$) by applying an external magnetic field. Our calculations reveal that the AHC component $\sigma_{xy}$ in the case of $M \parallel z$ is dramatically enlarged compared to the AHC component $\sigma_{yz}$ in the case of $M \parallel x$. Specifically, $\sigma_{xy}$ can reach a huge value of 1376 $\Omega^{-1}cm^{-1}$ when the Fermi level is located at 0.5 eV [Fig. 5(b)].

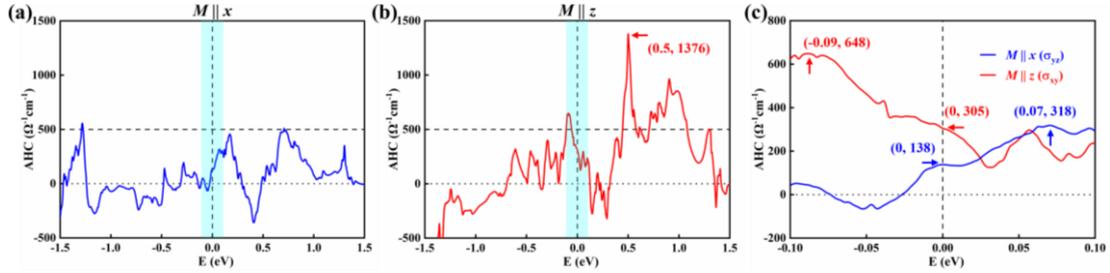

**Fig. 5.** The dependence of AHC on the position of the Fermi level. (a) AHC component $\sigma_{yz}$ in the case of $M \parallel x$. (b) AHC component $\sigma_{xy}$ in the case of $M \parallel z$. In (a) and (b), the Fermi level range from -0.1 to 0.1 eV is highlighted by the cyan areas. (c) The zoom-in AHC $\sigma_{yz}$ and $\sigma_{xy}$. In (b) and (c), the first and second numbers in the brackets are the values of the Fermi level and AHC, respectively.

For practical applications, the AHC near the Fermi level is crucial. We thus further study the AHC of $Cr_3Ge_2Te_6$ with the Fermi level ranging from -0.1 to 0.1 eV [see cyan areas in Fig. 5(a) and (b)]. Figure 5(c) shows the detailed dependence of AHCs on the Fermi level in this range. In the case of $M \parallel x$, $Cr_3Ge_2Te_6$ has an AHC of 138 $\Omega^{-1}cm^{-1}$ when the Fermi level is located at zero. We also find that the AHC almost monotonically increases to 318 $\Omega^{-1}cm^{-1}$ when the Fermi level is gradually tuned to 70 meV. In experiments, this could be realized through doping electrons into $Cr_3Ge_2Te_6$, e.g., the substitution of arsenic atoms for germanium atoms. In the case of $M \parallel z$, the AHC at the Fermi level is up to 305 $\Omega^{-1}cm^{-1}$. Surprisingly, the AHC can reach as large as 648 $\Omega^{-1}cm^{-1}$ when the Fermi level is located at -90 meV. Similarly, doping gallium atoms in the Ge sites of Cr$_3$Ge$_2$Te$_6$ should be appealing to realize such large AHE in experiments. The large AHC of the FM Cr$_3$Ge$_2$Te$_6$, which can be tuned by regulating its magnetization or doping, makes it very suitable for designing advanced spintronic devices.

Over the past decades, the development of highly sensitive and miniaturized magnetic devices has become more and more important. We thereby investigate the magnetic and electronic properties of Cr$_3$Ge$_2$Te$_6$ thin film with 5L thickness [Fig. 6(a)]. To determine the magnetic ground state and $T_C$ of 5L Cr$_3$Ge$_2$Te$_6$ thin film, we calculate its magnetic parameters based on the Hamiltonian Eq. (1). As listed in Table 1, the first nine NN Heisenberg exchange interactions of 5L Cr$_3$Ge$_2$Te$_6$ thin film are dominated by FM couplings whose values are much larger than those of AFM couplings. Due to these much stronger FM couplings, our Monte Carlo simulations reveal that 5L Cr$_3$Ge$_2$Te$_6$ thin film has a FM magnetic ground state with a $T_C$ of 405 K. Compared with the Cr$_3$Ge$_2$Te$_6$ bulk, 5L Cr$_3$Ge$_2$Te$_6$ thin film has a smaller MAE of 0.25 meV/ Cr. Hence, it should be easier to regulate the magnetic moment towards the hard axis (i.e., $M \parallel z$) by applying an external magnetic field to 5L Cr$_3$Ge$_2$Te$_6$ thin film.

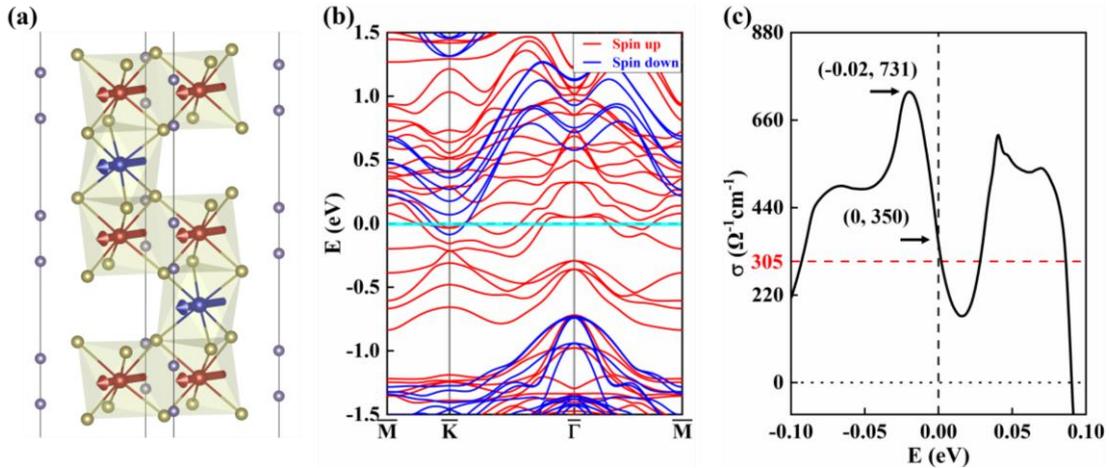

**Fig. 6.** Crystal and electronic properties of 5L Cr$_3$Ge$_2$Te$_6$ thin film. (a) The side view of the crystal structure. (b) Spin-resolved band structure without SOC. The energy interval highlighted by the cyan around the Fermi level is used to calculate spin polarization, $P$.

(c) AHC components, $\sigma_{xy}$, when the 5L $Cr_3Ge_2Te_6$ thin film has an out-of-plane magnetization. The AHC components $\sigma_{xy}$ of $Cr_3Ge_2Te_6$ bulk in the case of $M \parallel z$ with its Fermi level being located at zero is highlighted by red dash line. The first and second numbers in the brackets are the values of the Fermi level and AHC, respectively.

Figure 6(b) depicts the spin-resolved FM band structure of 5L $Cr_3Ge_2Te_6$ thin film. Only one band of minority states (i.e., spin down) around the high-symmetry point of $\overline{K}$ across the Fermi level, indicating that the 5L $Cr_3Ge_2Te_6$ thin film could have a higher spin polarization. Our calculation of the spin polarization from the integral total DOS of 18 meV near the Fermi level [see the cyan energy interval in Fig. 6(b)] indicates that the spin polarization of the 5L $Cr_3Ge_2Te_6$ thin film is as large as 99.8%. When the 5L $Cr_3Ge_2Te_6$ thin film has an out-of-plane magnetization, it has a greater AHC of $350\ \Omega^{-1}cm^{-1}$, compared with that ($305\ \Omega^{-1}cm^{-1}$) of $Cr_3Ge_2Te_6$ bulk [see Fig. 6(c)]. Surprisingly, the AHC of the 5L $Cr_3Ge_2Te_6$ thin film can reach up to $731\ \Omega^{-1}cm^{-1}$ when its Fermi level is located at -20 meV. This means it is easier to achieve larger AHE by doping holes into the 5L $Cr_3Ge_2Te_6$ thin film in experiments. Therefore, the nearly half-metallic ferromagnetism with the high $T_C$ above the room temperature and larger AHC enable the 5L $Cr_3Ge_2Te_6$ thin film greatly promising for designing advanced nano-scale electronic devices.

## 4. Conclusions

In conclusion, using the first-principles study and Monte Carlo simulations, we systematically investigate the magnetic and electronic properties of the self-intercalated vdW magnet, $Cr_3Ge_2Te_6$. Our calculations reveal that $Cr_3Ge_2Te_6$ is thermodynamically stable and that it is a room-temperature in-plane ferromagnet with a high $T_C$ of 492 K. By scrutinizing the electronic structure of the FM $Cr_3Ge_2Te_6$, we find that its spin polarization reaches 90.9%, indicating a nature of near half-metallicity in $Cr_3Ge_2Te_6$. Because of its ferromagnetism and the strong SOC effect, $Cr_3Ge_2Te_6$ exihibts AHCs of $138\ \Omega^{-1}cm^{-1}$ and $305\ \Omega^{-1}cm^{-1}$ when its magnetization is along the *x* and *z* axes, repsecitvely. Furthermore, these AHCs can be dramatically increased by doping holes or electrons in experiments. Lastly, the 5L $Cr_3Ge_2Te_6$ thin film also possesses in-plane ferromagnetism with a high $T_C$ above room temperature. More interestingly, the 5L $Cr_3Ge_2Te_6$ thin film shows better performance with a higher spin polarization and larger AHC, compared with $Cr_3Ge_2Te_6$ bulk. Our work demonstrates a novel self-intercalated $Cr_3Ge_2Te_6$, which has the room-temperature ferromagnetism, high spin polarization, large AHC, and may provide great opportunities for designing advanced nano-scale spintronic devices.

## Acknowledgements

This work was supported by the National Key R&D Program of China (Grant No. 2022YFA1403301), the National Natural Sciences Foundation of China (Grants No. 12104518, 92165204) and Guangdong Provincial Key Laboratory of Magnetoelectric Physics and Devices (No. 2022B1212010008). Yusheng Hou acknowledges the support from Fundamental Research Funds for the Central Universities, Sun Yat-Sen University (No. 24qnpy108).

## References

[1] Novoselov K S, Geim A K, Morozov S V, Jiang D-e, Zhang Y, Dubonos S V, Grigorieva I V and Firsov A A 2004 *science* **306** 666

[2] Duong D L, Yun S J and Lee Y H 2017 *ACS Nano* **11** 11803

[3] Lin Z, Peng Y, Wu B, Wang C, Luo Z and Yang J 2022 *Chinese Physics B* **31** 087506

[4] Zhang W, Wong P K J, Zhu R and Wee A T S 2019 *InfoMat* **1** 479

[5] Zhang X, Wang J, Zhu W, Zhang J, Li W, Zhang J and Wang K 2024 *Chinese Physics Letters* **41** 067503

[6] Zhu W, Xie S, Lin H, Zhang G, Wu H, Hu T, Wang Z, Zhang X, Xu J, Wang Y, Zheng Y, Yan F, Zhang J, Zhao L, Patané A, Zhang J, Chang H and Wang K 2022 *Chinese Physics Letters* **39** 128501

[7] Burch K S, Mandrus D and Park J-G 2018 *Nature* **563** 47

[8] Huang B, Clark G, Navarro-Moratalla E, Klein D R, Cheng R, Seyler K L, Zhong D, Schmidgall E, McGuire M A, Cobden D H, Yao W, Xiao D, Jarillo-Herrero P and Xu X 2017 *Nature* **546** 270

[9] Gong C, Li L, Li Z, Ji H, Stern A, Xia Y, Cao T, Bao W, Wang C, Wang Y, Qiu Z Q, Cava R J, Louie S G, Xia J and Zhang X 2017 *Nature* **546** 265

[10] Wang N, Tang H, Shi M, Zhang H, Zhuo W, Liu D, Meng F, Ma L, Ying J, Zou L, Sun Z and Chen X 2019 *Journal of the American Chemical Society* **141** 17166

[11] Wang Z, Zheng H, Chen A, Ma L, Hong S J, Rodriguez E E, Woehl T J, Shi S-F, Parker T and Ren S 2024 *ACS Nano* **18** 23058

[12] Yang R, Mei L, Lin Z, Fan Y, Lim J, Guo J, Liu Y, Shin H S, Voiry D, Lu Q, Li J and Zeng Z 2024 *Nature Reviews Chemistry* **8** 410

[13] Zhao X, Song P, Wang C, Riis-Jensen A C, Fu W, Deng Y, Wan D, Kang L, Ning S, Dan J, Venkatesan T, Liu Z, Zhou W, Thygesen K S, Luo X, Pennycook S J and Loh K P 2020 *Nature* **581** 171

[14] Coughlin A L, Xie D, Zhan X, Yao Y, Deng L, Hewa-Walpitage H, Bontke T, Chu C-W, Li Y, Wang J, Fertig H A and Zhang S 2021 *Nano Letters* **21** 9517

[15] Liu Y and Petrovic C 2018 *Physical Review B* **98** 195122

[16] Liu Y, Abeykoon M, Stavitski E, Attenkofer K and Petrovic C 2019 *Physical Review B* **100** 245114

[17] Tang B, Wang X, Han M, Xu X, Zhang Z, Zhu C, Cao X, Yang Y, Fu Q, Yang J, Li X, Gao W, Zhou J, Lin J and Liu Z 2022 *Nature Electronics* **5** 224

[18] Miao W-T, Zhen W-L, Lu Z, Wang H-N, Wang J, Niu Q and Tian M-L 2024 *Chinese Physics Letters* **41** 067501

[19] Zhao D, Zhang L, Malik I A, Liao M, Cui W, Cai X, Zheng C, Li L, Hu X, Zhang D, Zhang J, Chen X, Jiang W and Xue Q 2018 *Nano Research* **11** 3116

[20] Saha R, Meyerheim H L, Göbel B, Hazra B K, Deniz H, Mohseni K, Antonov V, Ernst A, Knyazev D, Bedoya-Pinto A, Mertig I and Parkin S S P 2022 *Nature Communications* **13** 3965

[21] Li B, Deng X, Shu W, Cheng X, Qian Q, Wan Z, Zhao B, Shen X, Wu R, Shi S, Zhang H, Zhang Z, Yang X, Zhang J, Zhong M, Xia Q, Li J, Liu Y, Liao L, Ye Y, Dai L, Peng Y, Li B and Duan X 2022 *Materials Today* **57** 66

[22] Zhang C, Liu C, Zhang J, Yuan Y, Wen Y, Li Y, Zheng D, Zhang Q, Hou Z, Yin G, Liu K, Peng Y and Zhang X-X 2023 *Advanced Materials* **35** 2205967

[23] Seo J, De C, Ha H, Lee J E, Park S, Park J, Skourski Y, Choi E S, Kim B, Cho G Y, Yeom H W, Cheong S-W, Kim J H, Yang B-J, Kim K and Kim J S 2021 *Nature* **599** 576

[24] Ni Y, Zhao H, Zhang Y, Hu B, Kimchi I and Cao G 2021 *Physical Review B* **103** L161105

[25] Ye F, Matsuda M, Morgan Z, Sherline T, Ni Y, Zhao H and Cao G 2022 *Physical Review B* **106** L180402

[26] Zhang Y, Ni Y, Zhao H, Hakani S, Ye F, DeLong L, Kimchi I and Cao G 2022 *Nature* **611** 467

[27] Wang J, Wang S, He X, Zhou Y, An C, Zhang M, Zhou Y, Han Y, Chen X, Zhou J and Yang Z 2022 *Physical Review B* **106** 045106

[28] Zhang Y, Lin L-F, Moreo A and Dagotto E 2023 *Physical Review B* **107** 054430

[29] Olmos R, Chang P-H, Mishra P, Zope R R, Baruah T, Liu Y, Petrovic C and Singamaneni S R 2023 *The Journal of Physical Chemistry C* **127** 10324

[30] Gu Y, Smith K A, Saha A, De C, Won C-j, Zhang Y, Lin L-F, Cheong S-W, Haule K, Ozerov M,

Birol T, Homes C, Dagotto E and Musfeldt J L 2024 *Nature Communications* **15** 8104

[31] Tan C, Deng M, Yang Y, An L, Ge W, Albarakati S, Panahandeh-Fard M, Partridge J, Culcer D, Lei B, Wu T, Zhu X, Tian M, Chen X, Wang R-Q and Wang L 2024 *Nano Letters* **24** 4158

[32] Susilo R A, Kwon C I, Lee Y, Salke N P, De C, Seo J, Kang B, Hemley R J, Dalladay-Simpson P, Wang Z, Kim D Y, Kim K, Cheong S-W, Yeom H W, Kim K H and Kim J S 2024 *Nature Communications* **15** 3998

[33] Huang C, Huo M, Huang X, Liu H, Li L, Zhang Z, Chen Z, Han Y, Chen L, Liang F, Dong H, Shen B, Sun H and Wang M 2024 *Physical Review B* **109** 205145

[34] Zhang Y, Ni Y, Schlottmann P, Nandkishore R, DeLong L E and Cao G 2024 *Nature Communications* **15** 3579

[35] Li D, Wang M, Li D and Zhou J 2024 *Physical Review B* **109** 155153

[36] Liu Y and Petrovic C 2018 *Physical Review B* **98** 064423

[37] Liu Y, Hu Z, Abeykoon M, Stavitski E, Attenkofer K, Bauer E D and Petrovic C 2021 *Physical Review B* **103** 245122

[38] Li Q, Cheng Y, Zhao D, Huang Y, Wan X and Zhou J 2023 *New Journal of Physics* **25** 103020

[39] Ran C, Mi X, Shen J, Wang H, Yang K, Liu Y, Wang G, Wang G, Shi Y, Wang A, Chai Y, Yang X, He M, Tong X and Zhou X 2023 *Physical Review B* **108** 125103

[40] Wu Q, Yin Q, Zhang S, Hu T, Wu D, Yue L, Li B, Xu S, Li R, Liu Q, Lei H, Dong T and Wang N 2024 *Advanced Optical Materials* **12** 2301863

[41] Yin Y, Wan X, Tu X and Wang D 2024 *Physical Review B* **110** 104410

[42] May A F, Liu Y, Calder S, Parker D S, Pandey T, Cakmak E, Cao H, Yan J and McGuire M A 2017 *Physical Review B* **95** 174440

[43] Qiao L, Barone P, Yang B, King P D C, Ren W and Picozzi S 2024 *Physical Chemistry Chemical Physics* **26** 8604

[44] Kresse G and Furthmüller J 1996 *Physical Review B* **54** 11169

[45] Perdew J P, Burke K and Ernzerhof M 1996 *Physical Review Letters* **77** 3865

[46] Blöchl P E 1994 *Physical Review B* **50** 17953

[47] Kresse G and Joubert D 1999 *Physical Review B* **59** 1758

[48] Liechtenstein A, Anisimov V I and Zaanen J 1995 *Physical Review B* **52** R5467

[49] Moulkhalwa H, Zaoui Y, Obodo K O, Belkadi A, Beldi L and Bouhafs B 2019 *Journal of Superconductivity and Novel Magnetism* **32** 635

[50] Hao Z, Li H, Zhang S, Li X, Lin G, Luo X, Sun Y, Liu Z and Wang Y 2018 *Science Bulletin* **63** 825

[51] Kresse G, Furthmüller J and Hafner J 1995 *Europhysics Letters* **32** 729

[52] Nosé S 1984 *The Journal of Chemical Physics* **81** 511

[53] Mazin I I 1999 *Physical Review Letters* **83** 1427

[54] Mostofi A A, Yates J R, Lee Y-S, Souza I, Vanderbilt D and Marzari N 2008 *Computer Physics Communications* **178** 685

[55] Wu Q, Zhang S, Song H-F, Troyer M and Soluyanov A A 2018 *Computer Physics Communications* **224** 405